\documentclass{PoS}
\usepackage{amsmath}
\usepackage{amssymb}
\title{Hadronization revisited: the dynamics behind hadro-chemical equilibrium}

\ShortTitle{Hadronization revisited}

\author{\speaker{R. Stock},
        University of Frankfurt}

 \abstract{
The multiplicities of hadronic species, from pions to omega hyperons
created both in elementary, and in highly relativistic
nucleus-nucleus collisions, are known to exhibit a pattern that is
very well reproduced within the framework of the statistical model -
as applied, in its canonical form, to the former class of reactions,
and in the grand canonical version to the latter. To understand the
origin(s) of this apparent equilibrium we revisit, first, the
hadronization models developed for $e^+e^-$ annihilation to hadrons
via partonic jet fragmentation that lead to a color pre-confinement
state, occuring at the end of the perturbative QCD DGLAP evolution,
which lends itself to color singlet formation that is then recast in
terms of massive singlet clusters, in a non perturbative model.
These clusters decay into the known hadron/resonance spectrum, the
eventual hadronic species yield distribution thus being seen as the
result of a perturbative evolution toward potential color singlet
''mass'' distributions that are subsequently acquiring excited non
perturbative hadronic condensate structures, that finally decay
statistically (under phase space governance) to on-shell
hadrons/resonances. Thus the success of a classical canonical
ensemble description is understood as the consequence of various
stochastic elements occuring in the dynamical evolution that ends
with hadronic freeze-out. Turning to hadronization in relativistic
nucleus-nucleus collisions we take advantage of the above singlet
cluster pre-hadronization picture. These clusters are spatially
isolated objects in the jet hadronization evolution, but with given
extended spatial size, of the order of 1 fm. Lacking a detailed
model of partonic transport evolution in highly relativistic A+A
collisions we assume that hadronization occurs from similar color
singlet clusters that will, however, overlap spatially owing to the
extreme overall energy density. An ensuing cluster overlap coherence
may be symbolically understood by means of a percolation model.
Cluster overlap increases with $\sqrt{s}$, A and collision
centrality. In the limit of an extended coherent volume,
hadronization is free of local quantum number constraints. After
super-cluster decay the yield distribution is captured by the grand
canonical version of the statistical equilibrium model which
features strangeness saturation (large volume limit).}

\FullConference{Critical Point and Onset of Deconfinement\\
         July 3-6 2006\\
         Florence, Italy}

\begin{document}

\section{Introduction}

The average multiplicity of hadronic species made up by the three
lightest quark flavours has been well characterized in a multitude
of elementary collision processes, e.g. $e^+e^-$, $pp$ and $p
\overline{p}$, and also in relativistic nucleus-nucleus collisions.
In the present article we will be first concerned with LEP $e^+e^-
\rightarrow$ hadron data at Z$_0$ energy ($\sqrt{s}$=92 GeV). The
ideas about hadron production dynamics in such collisions center on
a partonic shower evolution governed by perturbative QCD which ends
in color singlet formation \cite{1}, in the form of nonperturbative
QCD clusters \cite{2,3} or strings \cite{4}, which finally decay to
real hadrons and resonances \cite{3,5}. The most remarkable property
of the resulting multiplicity distributions, from pions to omega
hyperons, is that they are well described by a classical canonical
ensemble \cite{6,7}, i.e. by a state of statistical equilibrium.

\begin{figure}
\begin{center}
\includegraphics[width=7cm]{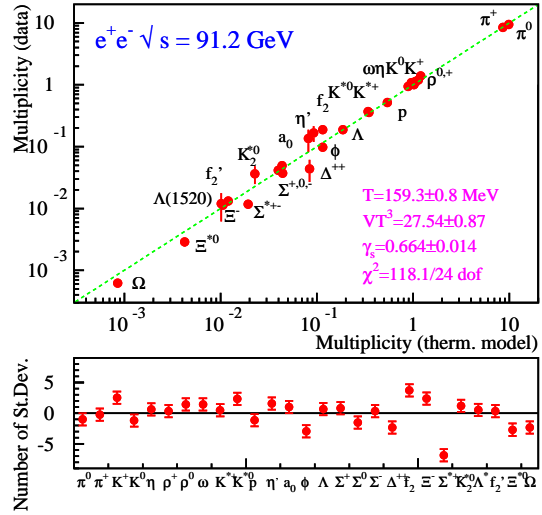}
\caption{Hadron production in $e^+e^-$ annihilation at
$\sqrt{s}$=91.2 GeV, with canonical model [6] fit.}
\end{center}
\end{figure}

Fig.1 illustrates the success of the statistical model description
[6] for LEP $e^+e^-$ annihilation to hadrons. This is a striking
outcome, considering the high energy that is initially distributed
over a small number of partons. Ever since Hagedorns first
systematic introduction of the statistical hadronization model
\cite{8} there has been fierce debate about how this apparent
equilibrium feature could be either dynamically acquired
(''thermalization'') or, alternatively, be a feature of quantum
mechanical decay of strings or clusters, occuring under dominance of
the phase space weights corresponding to the QCD hadronic/resonance
mass (spin etc.) spectrum. I.e. is the canonical ordering of
multiplicities essentially an ''acquired'' or a ''lost'' property?
The answer must be found mostly in the non perturbative phase
(recall, however, the ''pre-confinement'' idea of Amati and
Veneziano \cite{1,9}), a task that might become tractable given the
recent progress of npQCD hadron theory \cite{10}.

A remark of warning is in order here. The apparent canonical
multiplicity order has often inspired the statement that the final
hadronic state has ''maximum entropy'', in general. This is an
inaccurate statement as is obvious e.g. from LEP di-jet final states
which feature almost the opposite, near minimal entropy as far as
the pencil-sharp track topology signature is concerned.

\begin{figure}
\begin{center}
\includegraphics[width=7cm]{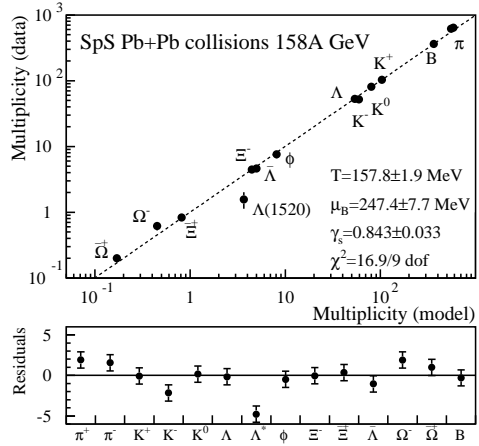}
\caption{Hadron production measured in $4 \pi$ by NA49 in central
Pb+Pb collisions at $\sqrt{s}$=17.3 GeV, compared to prediction [11]
by the grand canonical model.}
\end{center}
\end{figure}

We revisit hadronization in elementary collisions for the following
reason. It has turned out that the hadronization output from
relativistic nucleus-nucleus collisions, as studied at the AGS, SPS
and at RHIC, follows an analogous equilibrium ensemble pattern, this
time represented by the grand canonical (GC) statistical ensemble
\cite{11,12,13,14,15}. A typical example \cite{11} of a GC fit to
the hadronic multiplicities, from pions to omega hyperons as
observed by the SPS experiment NA49 in central Pb+Pb collisions at
$\sqrt{s}$ = 17.3 GeV is shown in Fig.2. The conceivable causes of
equilibrium are more diverse here. The observed thermal equilibrium
could be

\begin{itemize}
\item[a.] dynamical achieved by partonic thermalization prior to
hadronization \cite{16,17},
\item[b.] or be the outcome of stochastic and phase space effects
governing the decay process to hadrons/resonances \cite{18,19,20},
during hadronization,
\item[c.] or, finally, be the effect of hadron/resonance
inelastic rescattering cascades occuring in the dense medium right
after hadronization \cite{21,22} such that hadronic freeze-out does
not occur at, but after hadronization.
\end{itemize}

\noindent Note that, in option c, the observed grand canonical
temperature (T $\approx$ 160 MeV at top SPS and RHIC energy) should
be only a lower limit to the hadronization transition temperature.
The latter being an essential ingredient of the long-sought phase
diagram of QCD matter, and the clarification of the dynamical origin
of the grand canonical multiplicity pattern thus of crucial
importance concerning the significance of nuclear collision data.
Option a is of acute interest toward an explanation of the
''elliptic flow'' spatial emission pattern exhibited by the recent
RHIC data \cite{17} which appears to result from global hydrodynamic
parton flow anisotropies created in the early partonic phase of the
A+A dynamical evolution. Clearly, only option b describes the
elementary collisions which should thus be revisited.

The idea of this paper is thus to try out the following line of
argument:

\begin{enumerate}

\item Revisit the pQCD analysis of the truly elementary process
$e^+e^- \rightarrow$ di-jets $\rightarrow$ hadrons at LEP energy, in
particular the transition phase at the end of the perturbative
evolution leading to color singlet formation \cite{1}, a dynamical
pQCD process that gives rise to ''massive'' singlet clusters
\cite{2,3} which are recast into npQCD language ''pre-hadronic''
clusters. Interpreting them as a superposition of highly excited
hadronic resonance states their subsequent quantum mechanical decay
''gives birth'' \cite{5,8,18,19,20} to the canonical ensemble
population that is attested by the successful description within the
modern version \cite{6} of the Hagedorn model. This model also
provides for a description of hadronization in p+p collisions at SPS
energy \cite{7}.

\item
The latter observation suggests that the dynamical details of the
partonic pQCD phase are washed out during the subsequent npQCD
singlet cluster to on-shell hadron/resonance evolution. We note that
this is {\bf only} true for the multiplicity distribution of
hadronic species composed of the two light quark species. The
strangeness abundance appears to preserve memory of its initial
suppression, requiring a fudge factor.

\item This cluster hadronization model might, thus, be universal.
Its postulated equivalence to the hadronization scheme exploited in
the string model \cite{4,5} is, thus, reassuring but falls beyond
the scope of the present analysis. The model should thus be also
applicable to hadronization in highly relativistic A+A collisions,
e.g. from top SPS to RHIC energy, 17 $\le \sqrt{s} \le$ = 200 GeV,
where the assumption of a partonic primordial dynamics appears
equally plausible.

\item Such central collisions of A $\approx$ 200 nuclei provide for a
primordial reaction volume which is both extended far beyond typical
confinement and hadronization cluster size (both of order 1fm), and
featuring a very high average energy density, well above the
critical density of about 1 GeV/fm$^3$ at which lattice QCD predicts
the onset of deconfinement \cite{23}. The primordial dynamics may
thus establish a parton plasma state, as is indicated e.g. by RHIC
observations of jet quenching \cite{24} and elliptic flow formation
\cite{17}. Anyway the hadronic multiplicity distribution is
universally well described, again, by the statistical hadronization
model \cite{11,12,13,14,15} but now in its grand canonical form.

\item Finally: supposing that the hadronization process is essentially
insensitive to the details of the preceding partonic evolution
(point 2 above), the degree of partonic equilibration in a QGP might
be of minor relevance to the hadronization outcome (option a above).
However the high energy density, occuring in an extended volume, may
change the singlet cluster decay mode, from one that occurs isolated
in vacuum, in elementary collisions, to the decay of more extended
complexes of overlapping clusters which decay in a quantum
mechanically coherent manner. Formation of such ''super-clusters''
would be naively expected to increase with $\sqrt{s}$, nuclear
projectile mass A, and collisions centrality, in accord with the
observation that the transition from canonical to grand canonical
hadronization indeed occurs smoothly, within these variables
\cite{15,25,26}. The transition from small system size at
hadronization (elementary collisions), to large coherent system size
(A+A collisions) - with a decay to the final hadrons/resonances
occuring from spatially extended ''super-clusters'' - then appears
to be reflected in the concurrent transition from a canonical to a
grand canonical description. Note that it has been shown \cite{15}
that, in fact, the latter ensemble can be regarded as the large
coherent volume limit of the former, as far as production of hadrons
composed of the 3 light flavours is concerned.
\end{enumerate}

\noindent It follows that our above option c, the hypothesis of
equilibration through inelastic hadronic final state cascades
occuring after hadronization \cite{21,22}, is not required as the
principally novel mechanism governing A+A hadronic multiplicity
equilibrium - at least at very high $\sqrt{s}$ where hadrons should
form directly from a preceding partonic evolution. We note, however,
that anti-protons and anti-lambda hyperons may be subject to some
final state annihilation, not as much at top RHIC energy where
$\mu_B \rightarrow $ 0 as at top SPS energy where the final hadronic
cascade expansion is governed by a considerable net-baryon excess.
Furthermore, the eventually observed $\Phi$ multiplicity may deviate
from its primordial hadronization level, due to ''coalescence''
interaction with the high K$^+$+K$^-$  density prevailing during the
early cascade stage.

We also note that, toward the lower SPS and AGS energy domain, the
above overall model of hadronic freeze-out directly from a preceding
partonic phase becomes inapplicable as the freeze-out temperatures
obtained from grand canonical ensemble analysis \cite{11,12} fall
well below of the parton-hadron coexistence line in the [T,
$\mu_B$]-plane that has recently been established by an extension of
lattice QCD predictions to finite baryochemical potential \cite{27}.
This observation suggests the occurence of a finite interval in
expansion time of the fireball, spent between hadronization and
hadronic species freeze-out. This phase should indeed be subject to
the inelastic rescattering cascades considered in rfs. \cite{21,22},
our option c above.

In spite of such caveats I will concentrate here mainly on the
question of what can be learned for A+A collisions, by revisiting
the 1980s QCD jet hadronization models, developed e.g. for $e^+e^-$
annihilation \cite{1,2,3,4,5}.

\section{Hadronization in $e^+e^-$ annihilation}

What follows is a brief sketch of QCD models developed in the 1980s
period where first comprehensive data concerning hadron formation in
the, perhaps, most ''elementary'' QCD process of $e^+e^-$
annihilation to jets of hadrons became available. At typical LEP
collider energies, $\sqrt{s} \approx$ 100 GeV, this process is
reflected, in the end, by a back-to-back di-jet configuration
featuring about 20 hadrons. Its first step, $e^+e^-$ annihilation,
creates a virtual photon, labeled ''virtual'' as it contains the
total initial c.m. energy of the $e^+e^-$ pair but has zero momentum
in that frame. In view of the energy law of real particles, E$^2$ =
p$^2$+m$^2$, this situation with p$^2$=0 suggests the interpretation
that a state of ''virtual mass'' M$^2$=E$^2_{cm}$ has been created.
According to quantum mechanics its live-time is $\tau$ = 1/M within
which it decays to a quark - anti-quark pair. Next, over a period
characterized by time scales t such that $1/M < t < t_0$, the
primary partons develop into multi-parton cascades or showers by
multiple gluon bremsstrahlung, as illustrated [1] in Fig.3.

\begin{figure}
\begin{center}
\includegraphics[width=7cm]{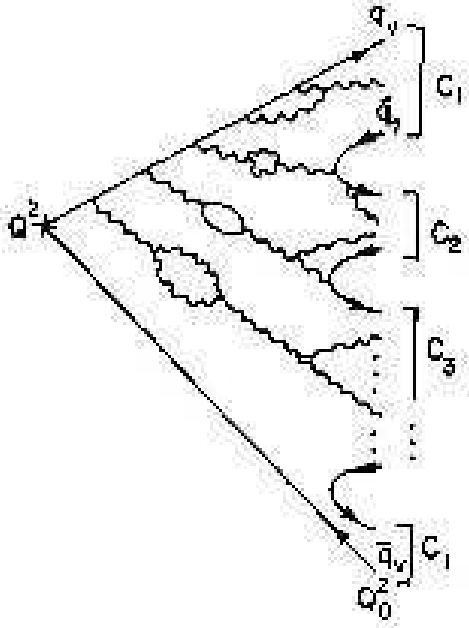}
\caption{DGLAP shower evolution of the primordial $q \overline{q}$
pair created in $e^+e^-$ annihilation, exhibiting singlet cluster
formation at the end [1].}
\end{center}
\end{figure}

These cascades, which tend to develop along the directions of the
primary partons owing to the ''collinear enhancement'' in QCD matrix
elements, are the precursors of the jets that are observed
experimentally. This rapid, hard process is described by the
so-called DGLAP-evolution of perturbative QCD; it leads from initial
high virtuality (and large momentum transfer-squared Q$^2$) to
partons of ''virtual mass'' 1/t$_0$ where t$_0$ approaches the
so-called QCD cutoff scale. It is defined by the requirement that
perturbative QCD is still applicable. After t$_0$ both Q$^2$ gets
lower and the time scale of the further evolution slower such that
the era of non perturbative QCD begins \cite{5}.

\begin{figure}
\begin{center}
\includegraphics[width=8.5cm]{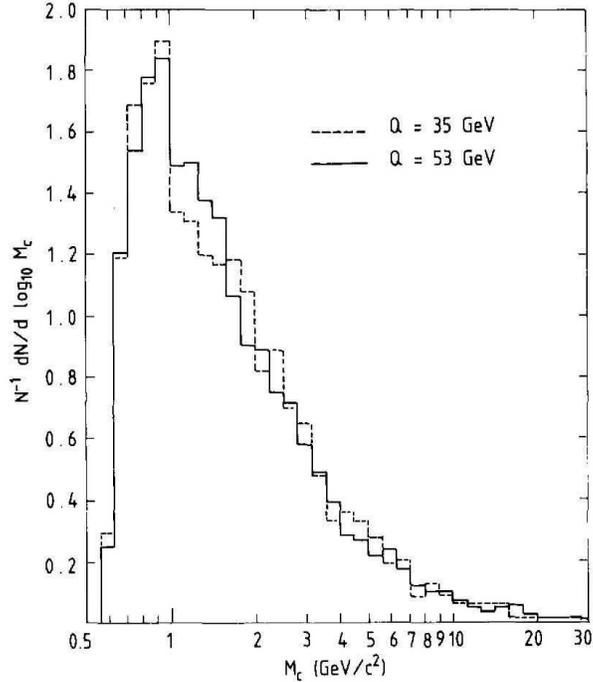}
\caption{Mass-distribution of color singlet clusters in $e^+e^-$
annihilation to hadrons [2].}
\end{center}
\end{figure}

However, right at the end of the pQCD phase the shower partons, i.e.
quarks and gluons, acquire a peculiar, ordered pattern in space and
color-flavour configuration that was called ''color
pre-confinement'' by its inventors, Amati and Veneziano \cite{1}.
Fig.3 indicates this ordering of the partons, occuring at t$_0$
through peculiarities of the pQCD matrix elements governing the
cascade sub-processes. The cascade thus breaks down into local
clusters $c_i$ where overall color neutralizes and ''hadron-like''
flavour combinations predominate; but all of this still in pQCD
vacuum. From the virtualities and momenta of the participating
partons one can construct the cluster c.m. systems and the (virtual)
cluster ''masses''. This task is performed, e.g. by the pQCD shower
Monte Carlo code ''HERWIG'' constructed by Webber and collaborators
\cite{2,5}. Fig.4 shows the resulting distribution of cluster masses
M$_c$, with mean value of about 1.5 GeV, and a steep fall-off toward
higher virtual masses. By construction under pQCD governance, these
clusters are color singlets, with typical dimension 1 fm. Obviously,
the stochastic elements of the overall pQCD shower evolution, acting
at the level of its various microscopic sub-processes, lead to a
rather broad probability distribution of cluster mass and flavour
content. Thus arises the first statistical influence on the path
toward hadronization.

With this singlet cluster stage we arrive at the end of perturbative
QCD applicability, entering non perturbative QCD territory for a
picture of the subsequent evolution. Nature, of course, does not
hesitate to proceed, which takes place via the parton-hadron phase
transition that ends with confined, on-shell hadrons. It is
remarkable, however, to see how smoothly the npQCD models of
hadronic structure match with the end product of the pQCD evolution,
e.g. dynamical color singlet formation [1]. In pQCD view these
clusters represent local ''excitations'' of the perturbative vacuum
as quantified by a virtual mass; pQCD evolution also suggests that,
toward and beyond the cutoff scale t$_0$, the time scale of the
evolution slows down considerably. It thus appears natural to assume
that typical npQCD structures, e.g. quark and gluon condensates,
replace the virtual vacuum excitation energy density of the pQCD
clusters, given sufficient time (now of order 1 fm/c) to develop. We
are now well beyond t$_0$. In the HERWIG simulation the cluster mass
spectrum of Fig.4 is then re-interpreted as a superposition of
excited hadronic resonances which subsequently decay quantum
mechanically, into the well-known spectrum of low lying on-shell
hadrons (and resonances) plus kinetic energy. This decay proceeds
under phase space governance, vis a vis the spectrum of
hadronic/resonance spins and masses, introducing the finally
observed yield distribution among the various hadronic species which
reflects their respective spin and phase space weights. Thus ends
the QCD evolution in the elementary process of $e^+e^-$ annihilation
to di-jets of hadrons. Their pencil-sharp, back to back correlation
in space (and in momentum space) clearly represents a highly
ordered, low entropy configuration, that results from the strictly
back to back configuration of the primordially produced $q
\overline{q}$ pair. The subsequent shower period and the eventual
decay into hadrons add modest stochastic transverse momentum
components, leading to the characteristic ''jet cone'' topology of
the finally observed hadrons. Remarkably, however, the yield
distribution among the final hadronic species (which arises from the
very same stochastic influences) presents itself as a maximum
entropy, equilibrium ensemble, as we infer from the success of the
statistical canonical ensemble fit to LEP jet hadronization data by
Becattini \cite{6}, shown in Fig.1.

{\it However:} in stressing the dynamical QCD origin of the apparent
hadronic species equilibrium (which has been the source of
inspiration for Hagedorns development of the statistical
hadronization model \cite{18} that has been endlessly challenged
until today) I have exaggerated the role of the stochastic elements
of the evolution, leading to loss of information. In fact the
canonical model fit shown in Fig.1 wraps up some crucial
information, concerning the conditions prevailing at the instant of
hadronic freeze-out to on-shell (in vacuum) hadrons and resonances,
in its {\it characteristic parameters}. Most notably, the derived
temperature amounts to T $\approx$ 160 MeV here, corresponding to an
energy density of about 0.7 GeV/fm$^3$. These values correspond
closely, on the one hand, to Hagedorns limiting conditions for a
hadronic system \cite{8} and, on the other hand, to modern lattice
QCD estimates of the parton-hadron phase transition temperature
\cite{23}. These conditions should, in fact, be universal to any
hadronizing system, irrespective of its dynamical pre-history. It is
thus reassuring to see that the same temperature governs
hadronization in central Pb+Pb collisions at $\sqrt{s}$ = 17.3 GeV,
as shown in Fig.2. We conclude that hadronic freeze-out in such
reactions occurs in the vicinity of the QCD parton to hadron
transition: a consistency check, concerning our above sketch of the
QCD evolution toward hadronization. All such high $\sqrt{s}$
hadronization data thus provide for an experimental determination of
the parton-hadron phase boundary.

The second parameter of the canonical model employed by Becattini
\cite{6} in Fig.1 is the total ''fireball'' volume contributing the
decay hadrons and resonances. In the reaction considered here,
$e^+e^-$ annihilation to hadrons at $\sqrt{s}$=91.2 GeV, it amounts
to 27 fm$^3$. This value compares well with expectations from the
cluster hadronization model by Webber et al. \cite{2,5} which, at
this energy, may feature about 10 clusters on average, of dimension
1 fm, decaying simultaneously but independent of each other. Note
that in jet hadronization the major fraction of the initial total
c.m. energy resides in hadron kinetic energy along the axis defined
by the initial $q \overline{q}$ pair.

Finally, the much debated ''strangeness under-saturation factor''
$\gamma_s$ is employed in the canonical fit shown in Fig.1. It
represents an additional fugacity factor, inserted for strangeness
production into the canonical partition functions, that takes
account of the apparent ''Wroblewski-suppression'' \cite{28} of
strangeness relative to light quark production which is
characteristic of all elementary collisions. At first sight $\gamma
_s$ may appear to be merely a fudge factor, within the framework of
the canonical model. It indicates that the canonical ensemble of
created hadrons deviates from global u, d, s flavour equilibrium,
leading to a relatively modest suppression of K and $\Lambda$
relative to pions but becoming increasingly effective toward
multiply strange hyperon production. Fig.1 shows that the canonical
fit is, in fact, critically constrained by the cascade and omega
hyperon data, thus clearly establishing the need of such an
auxiliary strangeness fugacity factor, $\gamma_s < 1$.

Auxiliary, as seen in the hadron/resonance population after
hadronization (with which the statistical model has to deal), this
apparent strangeness suppression may nevertheless be traced to the
pre-history of hadronization. In fact, the QCD evolution, sketched
above, consists chiefly of shower formation due to inelastic parton
interactions which, studied in isolation, always produce secondaries
under governance of strangeness suppression, basically due to the
higher strange quark mass which causes a corresponding penalty
factor. The emerging pre-hadronic clusters thus contain quarks with
a similarly suppressed strangeness fraction. Actually, a similar
population pattern arises in the alternative Lund string
hadronization picture \cite{4} where the flavour densities u:d:s are
predicted to be about 1:1:1/3. This relative under-population of
strangeness is preserved in the eventual cluster decay to on-shell
hadrons and hadronic resonances, which redistributes the initial
flavour content under governance of the relative phase space
weights, as implied by the hadronic mass and spin spectrum. The
canonical model fit, which takes a snapshot of the hadron/resonance
population right ''after birth'' [8,18,19], reflects the resulting
equilibrium distribution of yields, albeit within the restrictions
caused by initial strangeness under-population of the hadron
ensemble. Thus, $\gamma_s < 1$ reflects an absolute strangeness
non-equilibrium population which is, however, relatively distributed
in statistical equilibrium among the strange hadron species.

In summary we have, at first, stressed the remarkable feature that a
hadronic species equilibrium population can arise from fundamental
parton interactions which, in other respects, lead to final state
configurations (e.g. jets) that are dramatically remote from maximum
entropy. The reasons being found in stochastic elements governing
the dynamical QCD evolution \cite{1,2,3,4,5}. However, a more
detailed analysis of the key parameters of the canonical ensemble
model shows that they preserve critical information about the
dynamical pre-history of hadronization: the hadronization
temperature and energy density, the system volume prevailing at
hadronization, and the strangeness output arising from the preceding
microscopic QCD dynamics.

\section{Hadronization and hadronic freeze-out in A+A collisions}

In the previous chapter we have concentrated on hadron production in
the, perhaps, most elementary process of $e^+e^-$ annihilation to
jet hadrons because

\begin{enumerate}
\item The observed multiplicity order obeys the canonical
equilibrium pattern, initiated by Hagedorn, in the perhaps most
striking manner, due to the comprehensive LEP data illustrated in
Fig.1.
\item Furthermore, hadronization proceeds into vacuum, on-shell
products here, with no essential final state reconfiguration due to
inelastic hadronic interaction. Thus hadronization and hadronic
freeze-out occur in coincidence.
\item Finally, an elaborate dynamical QCD model exists which
features color pre-confinement singlet cluster formation and
non-perturbative hadronization, dominated by stochastic elements.
This model not only provides plausibility for the canonical ensemble
description of the hadronization outcome: it also helps to
understand the fundamental parameters of that model, e.g.
hadronization temperature, fireball volume, and strangeness
under-saturation.
\end{enumerate}

\noindent We can now deal with A+A collisions.

\hspace{1cm}

\subsection{Hadro-chemical equilibrium}

Turning to hadron formation in nucleus-nucleus collisions, from top
SPS to RHIC energy, $17.3\le \sqrt{s} \le 200$ GeV, Fig.2
illustrates the continuing success of a statistical hadronization
model description \cite{11} of NA49 data \cite{29}, gathered at
$\sqrt{s}=17.3$ GeV. {\it What is new:} the primary hadron
multiplicities require {\it a grand} canonical ensemble description,

\hspace{0.5cm}

 $<n_i>=\frac{(2J_i+1)}{(2\pi)^3} V \int d^3p \{exp
[(E_i+\mu_i)/T]\pm 1 \} ^{-1}$

\hspace{0.5cm}

\noindent where $E_i^2=p_i^2+m^2_i$ is the vacuum on-shell energy of
species i, and the penalty factor $exp(-E_i/T)$ is modified by the
global chemical potential $\mu_i$ for each species, thus taking care
of baryon number, strangeness and isospin conservation {\it on
average} over the entire fireball volume V. We see from Fig.2 that
the hadronic freeze-out temperature T=158 MeV is very close to the
value derived from canonical ensemble analysis of $e^+e^-
\rightarrow$ hadrons, Fig.1. It also agrees well with the recent
predictions of lattice QCD at finite baryo-chemical potential
\cite{27}, concerning the location of the parton-hadron coexistence
line in the [T, $\mu_B$] plane. As we shall show below the T,
$\mu_B$ values derived from grand canonical analysis \cite{12} of
STAR data obtained at top RHIC energy also fall close to this line.
One is, thus, tempted to conclude that hadronic species freeze-out
does very closely coincide with hadronization, both occuring at the
QCD phase boundary, which is thus {\it experimentally located from
top SPS energy to top RHIC energy.}

Furthermore we recall again that the hadrochemical freeze-out
temperature, observed here, agrees with the one observed in the
perhaps most simple and ''elementary'' QCD process of $e^+e^-$
annihilation to hadrons, where hadronization and hadronic freeze-out
coincide \cite{2,3}. One is, thus, tempted to suppose that
hadronization in central A+A collisions at such high $\sqrt{s}$
should (mutatis mutandis) resemble the QCD evolution sketched in the
previous chapter, i.e. shower multiplication of pQCD partons toward
decreasing virtuality resulting in a spatial color-flavour
correlation (''color pre-confinement'') \cite{1} that lends itself
to a re-interpretation in terms of non-perturbative QCD singlet
clusters whose initial virtual mass and spatial extension converts
to ''real'' hadronic mass and size. The clusters then decay to
on-shell hadrons and resonances under governance of phase space and
flavour conservation, thus producing the typical ''equilibrium''
population that is observed, both, in elementary and in A+A
collisions. This apparent equilibrium should thus, in both cases,
result from the stochastic elements governing the hadron formation
process: the hadrons are thus ''born into equilibrium''
\cite{8,18,19}. Right after birth the statistical ensembles provide
for a snapshot of these equilibrium distributions. This picture is
highly plausible for elementary collisions where the hadrons escape
almost instantaneously into vacuum, thus preserving the canonical
yield order. Not so in high $\sqrt{s}$ collisions of heavy nuclei.
The energy density corresponding to the grand canonical fit of Fig.2
amounts to about 0.6 GeV per fm$^3$ and, assuming a source radius of
7 fm, the source volume will double within about 2-3 fm/c, in a
simple isotropic fireball model. At first sight, such an early
expansion stage, still at considerable hadronic density, might
re-adjust the initial hadronic population ratios via inelastic
cascade processes. However such a dynamical cascade evolution of the
hadronic equilibrium distribution does, in fact, not occur here. We
{\it do} observe a freeze-out temperature of about 160 MeV, and {\it
not} the supposed ''relaxed'' temperature of about 135 MeV that
would (with $\epsilon \propto$ T$^4$) govern an equilibrium state at
twice the initial volume.

Thus we conclude that somewhat counter-intuitively, an initial
hadron/resonance ensemble at $\epsilon$ = 0.6 GeV/fm$^3$ and T=160
MeV stays unattenuated throughout the ensuing hadronic expansion
era. In fact it was shown by Bass and Dumitru \cite{30} that (at
RHIC energy) a yield distribution derived from hadronization,
supposed to occur at an even higher energy density ($\epsilon$=1
GeV/fm$^3$) survives the hadronic expansion era (modeled here by the
UrQMD transport model) essentially unattenuated. Noting that the QCD
parton to hadron phase transformation  occurs within the domain $0.6
< \epsilon < 1$ GeV/fm$^3$ according to recent lattice QCD,
hadronization and dynamical hadronic freeze-out should coincide
closely, in central A+A collisions from top SPS to RHIC energies. We
may, thus, indeed learn about matter at the phase boundary.

\subsection{Strangeness enhancement}

The fundamentally new, and unique feature  of central A+A collisions
is strangeness enhancement. More correctly we are dealing with the
fading away of strangeness suppression which is universally
characteristic of elementary collisions \cite{28,31}. One usually
quantifies the strange to light quark density ratio by the
''Wroblewski-coefficient'' \cite{28}

\hspace{0.5cm}

$\lambda_s = 2(s+ \overline{s})/(u+ \overline{u} +d + \overline{d})$

\hspace{0.5cm}

\noindent which amounts to about 0.25 in $pp, p \overline{p}$ and
$e^+e^-$ collisions \cite{31} but to about 0.50 in central A+A
collisions \cite{11}. This reflects in an increase of about two in
the production ratios $K^+/\pi$ and $\Lambda/\pi$, from p+p to
central Pb+Pb collisions at top SPS energy \cite{11,29}. Moreover,
suppression or enhancement feature a hierarchy in strangeness
number, i.e. the production ratios for $\Xi (s=2)$ and $\Omega
(s=3)$ relative to pions increase by about 6 and 15, respectively
\cite{15,32}. Note that these hyperons contain only a small fraction
of the total strangeness yield, due to their small production rate.
Their spectacular enhancements thus highlight an overall bulk
strangeness increase of about two. Nevertheless they dominate the
statistical model fits, due to their long lever arm, as is obvious
from Fig.1 and 2 which also illustrate the overall strangeness
enhancement pattern of central A+A collisions: from pions to omegas
the yield distribution drops down by a factor of about 0.5$\cdot
10^{-4}$ in $e^+e^-$ annihilation (Fig.1) but only by about
0.8$\cdot 10 ^{-3}$ (average of omega and anti-omega) in central
Pb+Pb collisions (Fig.2).

At the level of the statistical model description the hadron
multiplicity data for central heavy nucleus collisions are
universally well described by the grand canonical hadron/resonance
ensemble, from top AGS to top RHIC energies
\cite{11,12,13,14,15,18,19}. Fig.2 represents a typical example
\cite{11} obtained at top SPS energy; the NA49 data employed here
\cite{29} represent the total yields in $4\pi$ acceptance which,
even in a large acceptance detector system, require significant
extrapolation into unmeasured regions of phase space which are,
moreover, different for the various species, thus giving rise to a
species-dependent systematical uncertainty ranging from 10 to 30 \%.
In view of such experimental conditions the fit quality (with
$\chi^2 = 1.8/dof$) is quite satisfactory. This also supports the
significance of the notorious strangeness damping factor
$\gamma_s$=0.83, employed here: it suppresses the omega yields by a
factor of 0.56, well in excess of systematical errors. We will
return to second order corrections, such as this, in chapter 4. The
Wroblewski factor, corresponding to this fit, is $\lambda_s$=0.52.

\begin{figure}
\begin{center}
\includegraphics[width=9cm]{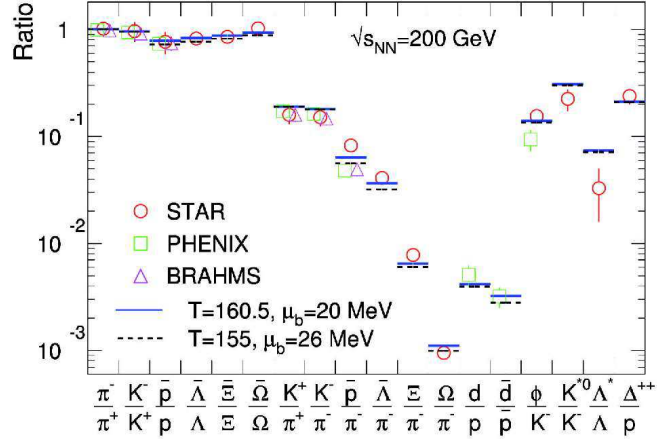}
\caption{Hadron production ratios at midrapidity in central Au+Au
collisions at RHIC, $\sqrt{s}$=200 GeV, and predictions by the grand
canonical model [12].}
\end{center}
\end{figure}

A grand canonical fit to recent RHIC data (central collisions of
Au+Au at $\sqrt{s}$=200 GeV) by Andronic, Braun-Munzinger and
Stachel \cite{12} is shown in Fig.5. These authors prefer to base
the grand canonical fit on hadron multiplicity ratios which are
determined at mid-rapidity owing to the limited acceptance of the
STAR and PHENIX experiments. We see, first of all, an omega to pion
ratio of $10^{-3}$ in close agreement with the NA49 data of Fig.2.
All anti-hadron to hadron ratios (for a given species) fall close to
unity, attesting to a small mid rapidity excess of quarks over
anti-quarks (as reflected in a small baryochemical potential, $\mu_B
\approx 20$ MeV) - in contrast to the situation encountered at
$\sqrt{s}$=17.3 GeV where $\mu_B \approx$ 250 MeV (Fig.2). Most
remarkably, the derived hadronization temperature, of T=160$\pm$5
MeV, agrees perfectly well with the value observed at
$\sqrt{s}$=17.3 GeV and, moreover with the one observed in $e^+e^-$
annihilation at $\sqrt{s}$=91.2 GeV (Fig.1). {\it This temperature
thus appears to be universal to hadronization at high $\sqrt{s}$}.
Furthermore it coincides with the ''critical'' temperature as
derived by lattice QCD \cite{22,27}, to prevail at the parton-hadron
phase boundary. Thus there is no evidence for a phase of statistical
dynamical inelastic equilibrium, occuring subsequent to
hadronization in A+A collisions \cite{21,22} - in $e^+e^-$
annihilation there is no such phase by definition. Hadronization and
hadro-chemical freeze-out thus occur in very close proximity at such
high $\sqrt{s}$. We note, however, that this simplicity is lost
toward the lower SPS, and notably the AGS energy regimes, where one
encounters grand canonical hadronic freeze-out temperatures as low
as 125 MeV \cite{11,12}, falling far below the QCD phase boundary
expected from recent lattice studies at a high baryo-chemical
potential \cite{27}. This indicates a qualitatively different
expansion dynamics \cite{21,22}, to which we return in sect. 4.3.

\subsection{Origin of strangeness enhancement}

According to our above argument strangeness enhancement in central
A+A collisions should be due to a mechanism located in the dynamics
prior to, or during hadronization, which removes (in part) the
strangeness suppression characteristic of elementary collisions.
This latter has oftentimes been associated with the conditions
prevailing in a small volume, i.e. both with the constraints imposed
by a strict, local conservation of quantum numbers, and with the
higher penalty factor resulting from the higher mass of the strange
- anti-strange flavour pair - both mechanisms together restricting
the accessible phase space for creation of secondary particles. In
the framework of hadronization, outlined in sec.2 for $e^+e^-$
annihilation to hadrons, this ''samll volume'' situation is captured
in the form of narrow local spatial clusters of quark-gluon
color-preconfinement which set the stage for hadronization. In the
Webber \cite{2} version of this model one expects to find about
10-15 such clusters at 50$\le \sqrt{s} \le$ 90 GeV, which are
isolated in vacuum and hadronize independent of each other. We note,
besides, that Becattini and Passaleva \cite{7} have shown that, for
a Lorentz invariant observable such as hadron multiplicity, these
hadronization subvolumes (which are boosted with respect to each
other) can be treated additively as a single, combined source. This
computational trick does, however, not affect the relative hadronic
composition which is still characteristic of the single cluster
quantum number constraints, prevailing in the non-overlapping
individual clusters.

However, now turning to A+A collisions at high $\sqrt{s}$, it is
obvious that their typical conditions, of a high energy and parton
collision frequency density \cite{26} occupying a large primordial
interaction volume, might lead to considerable spatial cluster
overlap. In fact under the naive expectation that, at RHIC energy,
$\sqrt{s}$=200 GeV, each interacting initial nucleon pair might
likewise create about 15 pre-hadronization clusters we would expect
a total of about 3000 such clusters in a central Au+Au collision.
The implicit assumption that the initially created partonic pQCD
showers still develop independent of each other, in this situation,
is probably inadequate, such that inter-shower multiplication of
secondaries will lead to a further increase of cluster spatial
density. Assuming a single cluster volume of about 1.5 fm$^3$
\cite{3} we would infer, from such naive arguments, that
non-overlapping clusters would occupy a total volume in excess of
5000 fm$^3$, much in excess of the expected total interaction volume
prevailing during the time interval of 2-3 fm/c, after the end of
pQCD shower evolution \cite{3}. We thus expect considerable cluster
overlap, to extended super-clusters: a large, quantum number
coherent volume is born, far in excess of confinement dimension. It
will develop toward hadrons under global, non-local quantum number
conservation, much like any other quantum mechanically coherent
object, and its decay will proceed via a QM decoherence transition.
I.e. its decay products will be in a quasi-classical state, the
''quasi'' reminding us that this state contains hadronic resonances
as well as on-shell hadrons \cite{8,17}. This hadron-resonance
''gas'' will occupy phase space uniformly (for a given total volume
and temperature) owing to the stochastic factors influencing both
the preceding QCD evolution, and the eventual decay. It is, thus,
not surprising that this state is well represented by a
quasi-classical grand canonical Gibbs ensemble \cite{18,19,20}. In
fact, the third fundamental parameter of this ensemble (besides T
and V) is the {\it global} chemical potential $\mu$ that represents
the fact that quantum number in the preceding decoherence decay was
preserved globally, i.e. only {\it on average} over the entire
volume. We wish to re-iterate that the GC ensemble, by itself,
merely captures a snapshot of the system right after its formation.
It has been the aim of the above line of argument to show that a
plausible QCD evolution can be conceived that leads, exactly, to the
conditions pictured in the statistical GC model. In fact, this model
provides for a satisfactory representation of the strangeness
enhancement phenomena.

The above picture, of pre-hadronization singlet cluster overlap in
A+A collisions reminds one of a percolation situation \cite{26}. In
fact, the degree of cluster overlap should increase with $\sqrt{s}$,
mass number A (i.e. with an exponent between 1/3 and 2/3) and
collisions centrality. The resulting expectation that, in general,
the Wroblewski ratio $\lambda_s$, and in more detail certain
characteristic hadronic yield ratios reflecting the relative
strangeness to non-strangeness output, should exhibit a smooth rise
(toward saturation at the GC level) within the above variables, has
been verified experimentally \cite{25,26,32,33,34,35}. Most
experiments have provided data for the dependence on centrality in
A$\approx$200 collisions \cite{32,33,34,35}, of hadronic production
ratios such as $K/\pi, \:\Lambda/\pi, \:\Xi/\pi$ and $\Omega/\pi$,
confronting them with a base-line of various elementary collision
yield ratios, extracted from $p+p, \: e^+e^-$ and $p+Be$ data at
similar energy. In addition NA49 has measured the $K/\pi,
\:\Lambda/\pi$ and $\Phi/\pi$ ratios in central collisions of light
nuclei, $^{12}C + ^{12}C$ and $^{28}Si + ^{28}Si$, at various SPS
energies \cite{25,32,34}. These data are usually displayed on a
common scale of ''participating' or ''wounded'' nucleon number as
derived, for each collision geometry, by Glauber simulation.

\begin{figure}
\begin{center}
\includegraphics[width=9cm]{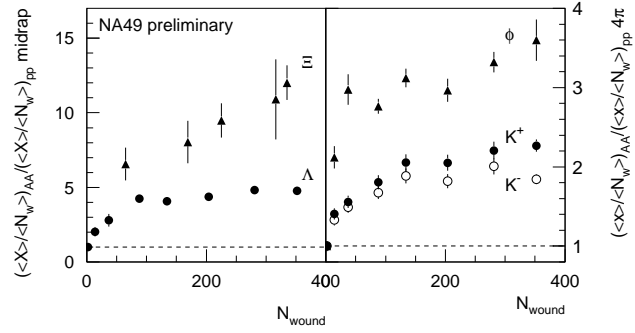}
\caption{Enhancement of the $\Lambda, \: \Xi, \: \Phi$ and $K$ yield
per wounded nucleon [25,34] as a function of collision centrality in
Pb+Pb at $\sqrt{s}$=17.3 GeV.}
\end{center}
\end{figure}

We illustrate typical examples of such studies in Fig.6 and 7. Fig.6
shows SPS results of NA49 \cite{24,33}, the left hand panel giving
the $\Lambda$ and $\Xi$ production rate per wounded nucleon in
various A+A collision geometries, as normalized to the corresponding
$p+p$ yields ($N_W$=2 by definition). The data shown here refer to
the respective hyperon yields at mid-rapidity. The right hand panel
shows the $K/W$ and $\Phi/W$ ratios vs. participant number, for the
same collisions and geometries, but it employs the extrapolation to
total $4\pi$ of the corresponding yields. Fig.7 (left panel) shows
similar data at top RHIC energy obtained by the STAR experiment
\cite{34} for the $\Lambda$ and $\Xi$ hyperons and their
antiparticle partners, in a plot similar to that of Fig.6 (left
panel), exhibiting the yields per participant in Au+Au collisions at
various centralities, as normalized to the corresponding $p+p$
yields (all taken at mid-rapidity).

We conclude that (irrespective of the different representations of
the above data) all strange or multi-strange hadronic species
exhibit a smooth rise (different in detail) of production yield
(relative to the strangeness suppression situation characterizing
small volume elementary collisions) with increasing interaction
volume as offered by the primordial reaction geometry. However, it
is important to note that it is not the ''large volume'' per se that
causes the transition from small volume, micro-canonical or
canonical strangeness suppression to grand canonical strangeness
saturation. Two large interacting clouds of a dilute nucleon gas
would, of course, yield exactly the $p+p$ particle ratios. It is the
coincidence of extended volume and high energy density, typical of
A+A collisions at high $\sqrt{s}$, that causes the effect. Such
conditions will, in fact, lead to an increasing overlap of QCD
pre-hadronization singlet clusters, as discussed above, thus
creating large, coherent sub-volume sections. Actually, this
description of the dynamical evolution may also imply an appropriate
model for ''Quark Gluon Plasma'' creation.

We see in Fig.6 that the $K$ and $\Lambda$ yields (which represent
the Wroblewski coefficient $\lambda_s$) ascend steeply, at first,
with system size but turn into saturation already for collision
systems of about 60-100 participants, corresponding to
semi-peripheral collision geometry in A=200 nuclear interactions.
For cascade hyperons saturation occurs at higher collision volume.
This behaviour would indeed follow from an increasing degree of
cluster overlap occuring with collision centrality, with percolation
super-cluster of increasing size being formed. The statistical model
also predicts a steep increase toward unity of the strangeness
suppression factor $\eta$, with increasing coherent volume
\cite{15,36}; the GC situation can thus be established as the large
volume limit of the canonical ensemble. In ref. \cite{15} is it also
shown that $\eta$ approaches unity (strangeness saturation) more
slowly with increasing hadronic strangeness. Thus, at first sight
there appears to be perfect overall harmony between data and
statistical model. However, a final step is missing. For s=1 one
obtains $\eta \rightarrow 1$ already at a remarkably small volume
\cite{15}, of about 60 fm$^3$, and even s=2 saturates at about 250
fm$^3$. Whereas the participant or wounded nucleon numbers where
these saturations appear to occur in top SPS to RHIC data are as
high as about 80 and 250, respectively, and the saturation
transition occurs more slowly with centrality. Now, 60 fm$^3$ is not
the ''fireball'' eigenvolume of about 80 interacting nucleons: the
relationship between experimental $N_{part}$ and statistical model V
is not straight forward.

\begin{figure}
\includegraphics[width=6.5cm]{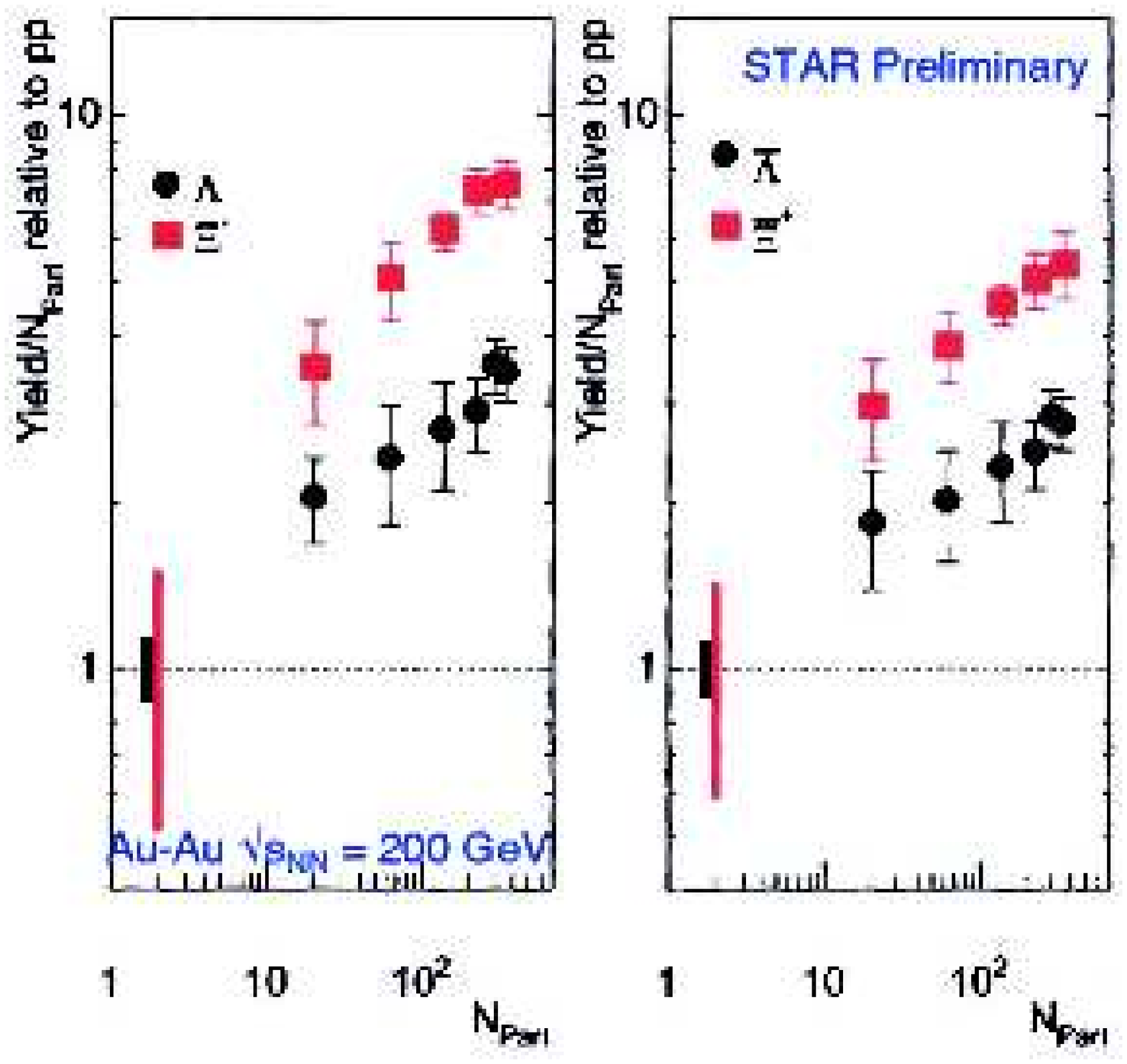}
\includegraphics[width=7.5cm]{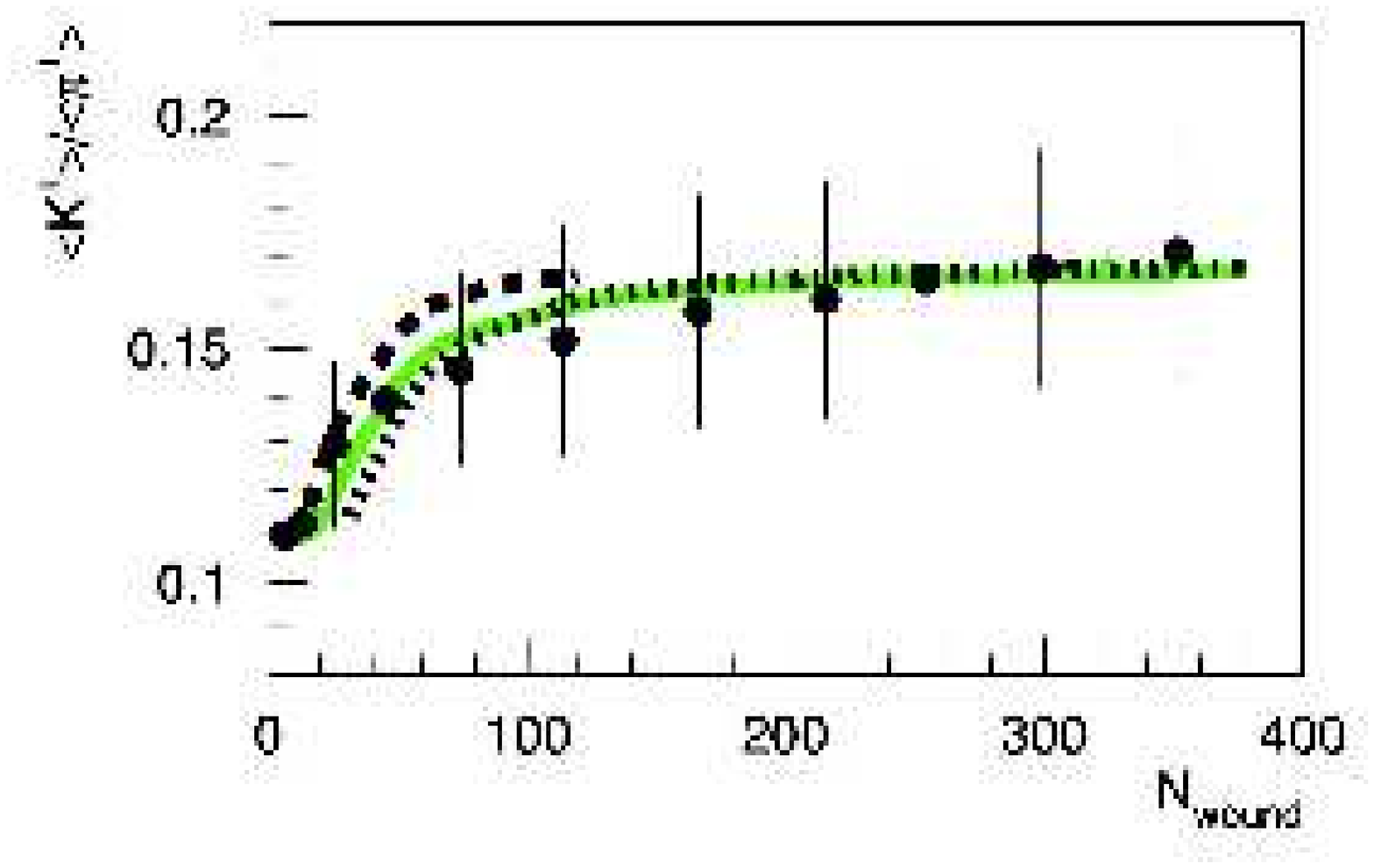}
\caption{$\Lambda$ and cascade hyperons and their antiparticles:
increase of the midrapidity yield per participant with centrality in
Au+Au at $\sqrt{s}$=200 GeV [32,35].}
\end{figure}

However, such a relationship can be established by means of a simple
percolation model \cite{26} which  quantifies the naive expectation
that super-cluster coalescence will develop gradually, with
increasing average energy density that should, by itself, also
exhibit a smooth relationship to collision centrality as monitored
by N$_{part}$. I.e. at first there will be several, relatively small
independent super-clusters hadronizing separately, which amalgamate
gradually with increasing centrality. Thus the participating nucleon
induced shower cascades, and their resulting local singlet clusters
will not, right onward from peripheral collisions, feed into a
single super-cluster with volume proportional to participant nucleon
number. Thus super-cluster size is not directly proportional to
N$_{part}$: it somehow lags behind. The percolation model (for
details see ref. \cite{26}) thus determines the super-cluster size
and number distribution, in bins of increasing centrality, then to
determine the appropriate strangeness suppression factors $\eta_i$
for each super-cluster $i$ according to the statistical model
\cite{15,36}. The final strangeness output results from a weighted
average  of all individual super-cluster contributions. A typical
result is shown in Fig.7 (right hand panel), which illustrates the
$K^+/\pi^+$ production ratio vs. $N_W$ as obtained by PHENIX
\cite{37} in Au+Au collisions at $\sqrt{s}$=200 GeV. The result of
the combined statistical and cluster percolation model gives a
perfect account of the relatively slow rise toward strangeness
saturation. Also included is a prediction for the Cu+Cu collisions
studied more recently at RHIC, which rises more steeply with N$_W$.
This illustrates the indirect relationship between N$_{part}$ or
N$_W$ and the corresponding size of super-clusters: at $N \approx $
100 the Au+Au collision system exhibits a typical peripheral
geometry, with extended contributions from dilute surface density
regions (in which participating nucleons create relatively small
energy densities) whereas the Cu+Cu collision is already near
central, thus featuring a higher average energy density.

We conclude that the overall body of strangeness enhancement
phenomena, typical of A+A collisions, can be understood in a picture
of amalgamating partonic singlet clusters that form at the end of a
preceding pQCD DGLAP shower evolution, at high $\sqrt{s}$, which we
have adapted here from previous QCD studies of jet hadronization in
the elementary $e^+e^-$ annihilation reaction \cite{1,2,3,4,5}. It
appears that the above line of argument, as suggested in the
introduction, can indeed establish a direct link between the
relatively straight forward QCD analysis of jet hadronization, and
the more complicated dynamics of A+A collisions which gives rise,
among other features specific to such collisions, to the phenomena
of strangeness saturation. Within this line of argument the success
of a statistical model description of the relative hadronic yield
patterns, i.e. the mysterious apparent equilibrium distribution, can
be also understood in detail, as a result of the preceding dynamical
evolution.

\section{Problems}

In the preceding chapters we have oftentimes ignored to face, in
detail, certain problems and objections, such as the extra fugacity
factor $\gamma_s$ or the problem of employing 4$\pi$ or midrapidity
data, etc.. We have thus given only an overall, qualitative
description of hadronization - but exactly this was our purpose. We
shall now first turn to the midrapidity problem which, as it turns
out, is partially related to the $\gamma_s$ question, and end with a
brief discussion of hadronization specifically at lower SPS and AGS
energies where our picture of an initial pQCD parton shower
evolution can not be expected to be valid.

\subsection{Midrapidity vs. 4$\pi$ data}

We start our consideration from two simple, limiting situations: at
very low energy, i.e. in the $\sqrt{s}$=1-2 GeV domain of the
Bevalac and SIS experiments, the entire rapidity distribution
resides in a gap $\Delta_y<2$. A single spherical fireball at a
temperature of 80 MeV spreads its particles over approximately
$\delta_y$=1.4 units, i.e. it essentially  fills the entire rapidity
gap, the experimentalists having a hard time removing, at least, the
spectator particles, but they will effortlessly cover the rather
narrow $\delta_y$ interval. Thus, essentially all existing data are
extrapolations to 4$\pi$, and smaller intervals are difficult even
technically because of the ''banana-shaped'' typical fixed target
spectrometer acceptances in the plane ($y, p_T$). On the contrary,
at extremely high energy (i.e. perhaps at the LHC energy
$\sqrt{s}$=5.5 TeV) the $y$-distribution becomes boost invariant
over about 10 units of rapidity, such that the surface ''corona''
contribution (from single collisions in the Woods-Saxon surface)
which will still reside in the vicinity of $\mid Y_{beam} \mid$
becomes a very small fraction of the total cross section. With
essentially flat rapidity distributions the acceptance does not
matter much, and one will concentrate on midrapidity data where all
detector acceptance overlap.

Unfortunately, however, all data discussed at present, from AGS to
RHIC energy, fall unhappily in between these extremes. From AGS to
top SPS energies the rapidity gap widens to $\Delta y$=3-6, too wide
to be described as the decay outcome of a single central fireball as
the interaction volume gets stretched by preponderance of
longitudinal over transverse expansion flow. Moreover both here, and
up to top RHIC energy, the rapidity distributions differ for the
various hadronic species, and are far from boost invariance flatness
\cite{38}. Thus, midrapidity hadron yield ratios are systematically
different from $4 \pi$ ratios. An example of this difficulty is
given by the SPS data shown in Fig.6. The left hand panel
illustrates the midrapidity ratio (per wounded nucleon) of the
$\Lambda$ yield in Pb+Pb collisions, relative to the one observed
(again at midrapidity) in p+p collisions at the same energy. Noting
that the $\Lambda$ distribution features even a minimum at
midrapidity for the latter system, it is no surprise to see the
yield ratio ascending to a value of about 4.5 with increasing
centrality. Whereas the $K^+$ data in $4 \pi$ geometry (right panel)
exhibit an increase with Pb+Pb collision centrality, of only a
factor of about two, over the p+p reference point that also refers
to $4 \pi$ extrapolated data here. As the $\Lambda$ and $K^+$ yields
both represent the major fractions of the s and anti-s production
rates, respectively, and should thus reveal a similar bulk
strangeness grand canonical saturation pattern, we are faced with a
serious apparent inconsistency. We have disregarded this in ch.3
because there our emphasis was to demonstrate, first of all, the
basic phenomena of relative strangeness increase with system size.

Before rushing to conclusions let us analyze the origin of the above
difficulties which have resulted in a vivid but premature
controversy among the various schools of statistical model
proponents \cite{11,12}. To this end we note, first, that the origin
of the above, and other apparent incongruencies resides in the fact
that {\it two} closely interwined mechanisms govern hadron
formation, in proceeding from elementary p+p (or p+Be \cite{32}) to
central A+A collisions. These mechanisms are well separated on the
overall time scale. Primordial baryons, and their net baryon number
content, are disentangled in the course of initial pQCD interaction.
In dependence on A, $\sqrt{s}$ and reaction geometry a specific
re-distribution arises in longitudinal phase space, {\it both} of
resulting net baryon number and energy density. This distribution
depends on the characteristic ''stopping power'', as offered by the
longitudinal thickness of the reactants. The higher the stopping
power, the wider the shift of the incident valence quarks, away from
initial $y_{proj}$ and $y_{targ}$ and toward midrapidity - alongside
with the buildup of high energy density.

At later times the clusters or super-clusters that approach
hadronization are differently composed according to their location
in $y$ space. Thus, for example at top SPS energy, there arises a
hierarchy of net baryon $y$ distribution shapes: from maximally
forward/backward peaked in minimum bias p+p collisions to near
''flat top'' shape in central Pb+Pb. This reflects in the
y-distribution of the (net) proton, $\Lambda, \: \Xi$ and $K^+$
yields  which tend to be much broader than those of the
corresponding anti-hadrons which are free of initial u, d quarks.
The result: evaluated in small bins of rapidity all hadron ratios
depend on the local rapidity. However we need to recall here that
the smallest bin size suitable for statistical model analysis is
given by the spread in rapidity resulting from the decay of an ideal
single, isolated fireball which, at a hadronization temperature of
160 MeV, amounts to $\Delta_y \approx 2$ \cite{11}. The relatively
narrow rapidity intervals, populated at AGS and lower SPS energies,
thus do not allow for an analysis in several separate bins, more
evidently so as we have to recall that the physics of equal mass
target and projectile collisions exhibits reflection symmetry about
midrapidity. One is thus left with only two options, to analyze
either in the interval ($y_{mid} \pm 0.5$), or in $4\pi$ , both
options not ideal - thus the vivid controversy concerning a correct
statistical model approach \cite{11,12}. We shall illustrate this
situation in the next section, devoted to the $\gamma_s$ problem,
but conclude, in the meanwhile, that statistical model analysis
retains at present a certain approximative character at intermediate
$\sqrt{s}$ \cite{11,13}.

\begin{figure}
\begin{center}
\includegraphics[width=7.5cm]{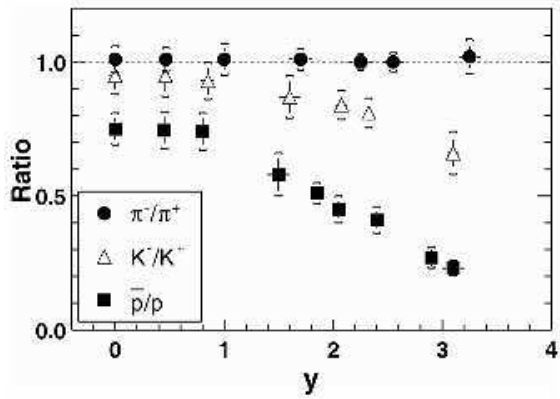}
\caption{Hadron production ratios vs. rapidity, central Au+Au
collisions at $\sqrt{s}$=200 GeV, BRAHMS experiment [38].}
\end{center}
\end{figure}

However, at top RHIC energy the wider rapidity gap, $\Delta_y
\approx 10$, (or even its half, because of reflection symmetry)
permits a statistical model analysis which is differential in $y$.
This fascinating new aspect has been discussed by R\"ohrich
\cite{39}, based on wide acceptance Au+Au collision data gathered by
the BRAHMS detector \cite{38}, which are shown in Fig.8.
Illustrating the remarks above concerning the stopping power effect
on the net baryon number rapidity distribution, the figure shows a
dramatic dependence of the antiproton to proton ratio: a short
''plateau'' region (a la boost invariance) governs the interval
$y_{CM} \pm 1$; this is also true for all other hadron yields and
ratios, measured here and by the PHOBOS and STAR experiments at
RHIC. But at $y_{CM} > 1$ the $\overline{p}/p$ ratio drops down
steeply toward 0.2 at $y \approx 3.5$, thus making close contact to
the top SPS energy values obtained {\it at midrapidity} by NA49
\cite{40}. The $K^-/K^+$ ratio follows a similar drop-off pattern,
to about 0.65, again matching closely with the top SPS energy value,
of about 0.6  \cite{29}. The deviation from unity of these ratios
reflects the density on initial valence u, d quarks, relative to the
density of newly created light and strange quark - anti-quark pairs,
i.e. the net baryon number density and its rapidity distribution.
Exactly this density ratio defines the baryo-chemical potential
parameter of the GC ensemble. Thus, in analyzing successive bins of
these rapidity distributions, the major variation in the GC fit
concerns the baryo-chemical potential $\mu_B$ which increases from
about 20 MeV (see Fig.5) at midrapidity, to about 150 MeV at $y \ge
3$ \cite{39}, while the hadronization temperature stays near
constant, at T=160 $\pm$ 5 MeV.

Referring to \cite{38,39} for detail we conclude, first, that
statistical model analysis can be carried out differentially in $y$
at top RHIC energy. Remarkably, the total rapidity interval is just
wide enough to permit such an analysis, but it is narrow enough, on
the other hand, to establish a situation which is quite different
from ideal boost invariance; such that local hadron composition
varies with the site of hadronization, in longitudinal phase space.

We are thus on firmer grounds, concerning statistical model analysis
at RHIC energy and beyond. Moreover, the above RHIC observations are
highly relevant for the general topic of hadronization in A+A
collisions at high $\sqrt{s}$. We see that hadronization occurs
''late'', and not from a single, globally coherent fireball volume
but from a sequence of super-clusters with limited internal rapidity
spread, and with a valence quark density which changes along the y
axis according to the primordial stopping process. We have shown in
sect.3.4 that grand canonical strangeness saturation does not
require all 400 participant nucleons to gather in a single
super-cluster (Fig.6), and ref. \cite{15} suggests that
$N_{part}$=40-80 suffices, even for s=3. It will be interesting to
see how these phenomena evolve at LHC energy.

One disturbing feature remains in the overall picture. The STAR data
\cite{35} illustrated in Fig.7 (left frame) do not really exhibit
the saturation pattern predicted by the model \cite{15}. The
hyperons yields per participant keep increasing all the way up to
fully central collisions, and this is also true for the SPS NA49
cascade hyperon yield \cite{34} shown in Fig.6 (left frame), whereas
the s=1 yields do indeed exhibit a saturation pattern. Could this
imply that full strangeness saturation of s=2 and 3 hyperons does
not really occur? In the light of the above discussion of a
hadronization situation with successive super-clusters, ordered in
rapidity and decaying each by themselves, we might be tempted to
conclude that they stay too small, in such a situation of extreme
longitudinal expansion flow. The ongoing analysis of RHIC data at
$\sqrt{s}$=64 GeV might clarify this. Or should we propose \cite{41}
that the extra suppression factor $\gamma_s$ equals unity only in
fully central collisions at RHIC, dropping down with smaller
participant number? We are apparently exchanging one evil by the
other - to which we shall turn next.

\subsection{The strangeness suppression factor Gamma}

The question whether an extra fugacity factor $\gamma_s$ suppressing
strange hadronic species (which is familiar from canonical model
studies of elementary collisions) is required in the GC approach is
still open. Furthermore even if this question is to be answered
affirmatively the origin of a $\gamma_s<1$ is not uniquely
understood \cite{11,12,14}. Let us first try some simple
considerations. The problem could be an off-shot of the above $4\pi$
vs. midrapidity discussion. We note that an ideal single fireball at
T=160 MeV creates a much broader rapidity distribution for $\pi$
than for $\Omega$, the reason being the decrease with hadron mass of
the average thermal velocity, of which $y$ represents the
longitudinal component. Thus $\sigma (y)$ of a Gaussian
parametrization decreases by more than a factor of two. Thus, in a
relatively narrow window centered at midrapidity one records a
higher fraction of the total $\Omega$ yield, as compared to the pion
yield fraction. I.e. the $\Omega/\pi$ ratio is higher at midrapidity
than in $4\pi$ perhaps by a factor of two \cite{11} for a
midrapidity window of $\Delta y=1$. To a somewhat lesser extent this
repeats for the cascade, $\Lambda$ and $\Phi$ to pion ratios. We
thus get a spurious extra strangeness enhancement, increasing toward
s=2 and 3, with midrapidity data, at the SPS (narrow gap). However
this argument is an oversimplification as hadrons are not produced
from an ideal single isotropic fireball even at SPS energy.

\begin{figure}
\begin{center}
\includegraphics[width=8cm]{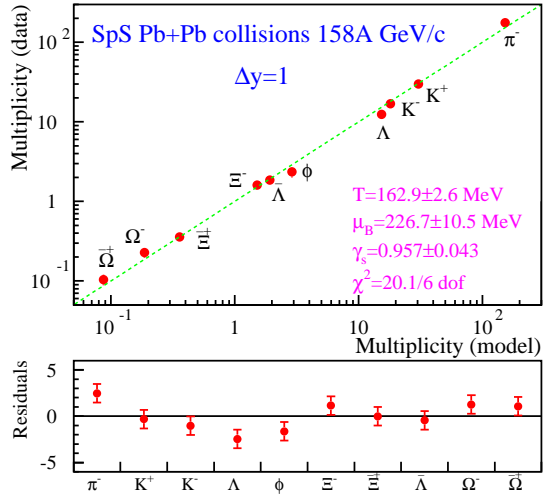}
\caption{Midrapidity hadron multiplicities in central Pb+Pb
collisions at $\sqrt{s}$=17.2 GeV [25,29] vs. grand canonical model
prediction by Becattini [11,42].}
\end{center}
\end{figure}

Nevertheless, this effect is indeed observed. Fig.9 shows a GC fit
by Becattini \cite{42} employing a $\Delta y=1$ midrapidity cut on
the NA49 data that result \cite{11} in the fit of Fig.2, in their
$4\pi$ version. The latter fit requires $\gamma_s=0.83$ with
adequate statistical significance whereas the midrapidity fit
procedure gives $\gamma_s=0.96 \pm 0.04$, compatible with unity, and
in agreement with the parallel midrapidity analysis of
Braun-Munzinger, Stachel and collaborators \cite{12} which employs
$\gamma_s=1$, throughout. We have shown their RHIC data analysis in
Fig.5. Comparing Figs.2 and 9 we see that the freeze-out
temperature, T=(160 $\pm$ 3) MeV, is insensitive to the alternative
choices, whereas the derived value of the baryo-chemical potential
$\mu_B$ is significantly lower at midrapidity than in $4 \pi$. This
reminds us of the stopping power effects discussed in sec.4.1.
Already at top SPS energy the final rapidity distribution of the
initial u, d valence quarks from the target and projectile nuclei
(which defines the net baryon number distribution) exhibits a slight
minimum at midrapidity, as is reflected directly in the
corresponding net proton and net $\Lambda$ distributions. As a
result the y-distributions of p and $\Lambda$ are drastically
different from the corresponding anti-hadron distributions, which
are Gaussian about midrapidity. To a lesser degree this also causes
the y distributions of all other valence quark carrying hadrons
(notably $K^+$ and $\Xi$) to be broader than those of their
respective anti-hadron partners. Again, it is clear that a
relatively narrow midrapidity cut will lead to hadron production
ratios which are different from the corresponding $4 \pi$ values. To
assess the possible consequences of this phenomenon Becattini
\cite{42} has formerly performed a GC analysis of a subset of then
{\it preliminary}  NA49 Pb+Pb central collision data, only retaining
hadrons that consist entirely of newly created quarks. The
corresponding fit is shown in Fig.10 and, within the limitations of
a poor statistical significance, we take note that $\gamma_s$ again
results at unity whereas the temperature stays unchanged throughout
the sequence of Figs.2,9 and 10. We note that the preliminary NA49
data, employed here, have been since corrected (lowering the
$\overline{p}$ and $\Omega$ yields to the final values employed in
Figs.2 and 9 which would improve the significance). This exercise
has thus to be repeated, and extended toward the lower RHIC
energies. We have illustrated it here to register the interesting
idea.

\begin{figure}
\begin{center}
\includegraphics[width=7cm]{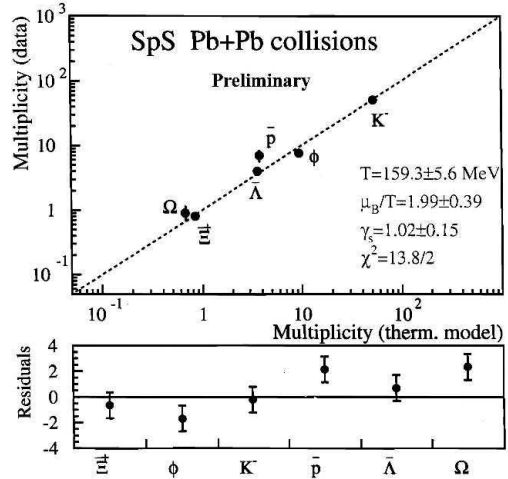}
\caption{Preliminary grand canonical model fit to central Pb+Pb
collision data in 4$\pi$, at $\sqrt{s}$=17.3 GeV, where only
valence-quark-free hadrons are included [42].}
\end{center}
\end{figure}

In conclusion it appears that the principal difficulties, discussed
in this and in the preceding section, might prohibit an ideal,
carefree definition for the proper employment of GC analysis at
intermediate $\sqrt{s}$. The $\gamma_s$ problem can, thus, not find
a satisfactory answer as of yet. This situation will change with the
advent of LHC data where we can expect to observe more nearly boost
invariant rapidity distributions.

\subsection{Grand Canonical analysis at
lower $\sqrt{s}$}

We have seen in sect.4.1 and 4.2 that the simple GC multiplicity
analysis with global parameters resides on an idealization of the
collision which cannot {\it exactly} fit physical reality, and
discrepancies are thus to be expected. Thus a new mechanism or a
modification of the basic scheme proves to be relevant only if it
leads to a major improvement of agreement  with the data. This
refers, among others, to the question whether a determination of the
(partonic phase) strange to non-strange density ratio {\it before}
hadronization is possible, and reflected in the $\gamma_s$ factor.
This would take into account that at high $\sqrt{s}$ this ratio
might be smaller in a quark gluon phase than in an equilibrium GC
hadronic ensemble. This expectation remains unverified  as of yet,
and likewise the proposal \cite{43} that hadronic freeze-out occurs,
not to a quasi-classical but to a mean field situation which implies
in-medium modified hadron masses. Besides being somewhat
counterintuitive (after freeze-out the hadrons should be on-shell)
this proposal introduces numerous further parameters but with a
modest net gain in fit significance.

However, the fit variations in Figs.2,5,9,10 do indeed illustrate
the general observation that the implied freeze-out temperature
stays rather insensitive to such second-order modifications,
contrary to the claims made in ref. \cite{12}. In fact, from top SPS
energy $\sqrt{s}$=17.3 GeV via RHIC at 64, 130 and 200 GeV the
temperature is compatible with a constant, global average value of
165$\pm 5$ MeV \cite{11,12,38}. This value does remarkably well
coincidence with recent lattice QCD predictions of the critical QCD
temperature, both at $\mu_B=0$ \cite{23} and at finite $\mu_B$
\cite{27}. Toward lower $\sqrt{s}$, the derived freeze-out
temperatures drop down smoothly until SIS energy, permitting an
interpolating fit \cite{11,12,44,45}. The situation is illustrated
in Fig.11, in the [$\mu_B, \:T$] plane. The GC freeze-out points are
seen to drop well below the phase coexistence line conjectured by
lattice QCD \cite{27} which, furthermore, predicts a critical point
of QCD to occur in the vicinity of $\mu_B$=300 MeV. This leads  to
important conclusions, firstly that the existence {\it and} location
of a critical point (and thus a first order phase transition region
toward higher $\mu_B$) may create critical fluctuations of the
hadronic composition \cite{46} in the lower domain of SPS energies,
also affecting hadronization and the resulting hadronic ratios. This
would lead to deviations from standard GC behaviour \cite{47}, which
are indeed observed \cite{48}.

\begin{figure}
\begin{center}
\includegraphics[width=8cm]{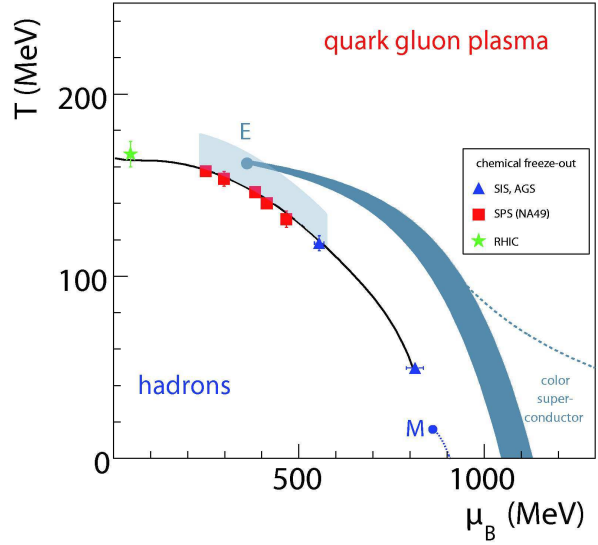}
\caption{QCD matter phase diagram illustrating hadronic freeze-out
points [11,12,14,36,43] and the parton-hadron phase boundary, also
indicating the expected critical point [27].}
\end{center}
\end{figure}

The second implication of Fig.11 is even more directly relevant for
the arguments discussed in the present article: that hadronic
freeze-out occurs later than hadronization, below $\sqrt{s}
\approx10$ GeV. It thus does {\it not} occur from a preceding
partonic cluster phase but from a dense hadronic medium, with mean
chiral fields \cite{43} and Goldstone bosons \cite{21} in place,
leading to short relaxation times of the participating hadrons. New
mechanisms for inelastic relaxation or equilibration could dominate
this medium directly at the coexistence line, such as the inverse,
by detailed balance, of string or cluster decays to many hadrons
\cite{22}, i.e. reactions such as

\hspace{3cm}  n pions  $\leftrightarrow X \leftrightarrow \Omega
\overline{\Omega}$

\noindent with n up to 10 as encouraged by the high hadron density
$\varrho$, which accelerates such processes in proportion to
$\varrho ^n$. It is thus conceivable that the system expands while
maintaining chemical equilibrium until the freeze-out temperature is
reached. This equilibrium can be maintained if the main relaxation
time constants are smaller than the expansion time scale (for
example the volume doubling time). At higher $\sqrt{s}$ the system
arrives at the phase boundary, from above, while its expansion flow
velocity fields are already fully developed in the preceding
partonic era \cite{17}, i.e. it ''races'' through the phase boundary
at an expansion time scale of about 2fm/c. Hadronic ''classical''
rescattering relaxation time constants are not conceivable to fall
down to,or below hadron diameter (here we disagree with
ref.\cite{21}), i.e. they can not really be much smaller than 2 fm/c
with the consequence that the system freezes out in the close
vicinity of the phase coexistence line. However, at low $\sqrt{s}$
there does not necessarily occur an extended partonic era (or non at
all). The system reaches something like a classical turning point
\cite{46} between compression and re-expansion, where dynamical time
scales are rather long. This turning point may, in fact, closely
coincidence with the conjectured critical point, at low SPS energies
(i.e. the system does not penetrate deeply into the partonic phase),
or it may even fall below the phase boundary, all together, so that
the system stays hadronic. But in either case there will occur a
much slower initial re-expansion here, {\it coinciding} with maximal
hadron and energy density. Thus $\tau_{relax} \ll \tau_{exp}$, and
chemical equilibrium can in fact be maintained for a while, such
that $T_{freeze-out} < T_c$. Indeed, $\tau_{exp}$ increases with
falling $\sqrt{s}$, as is required by Fig.11. This indicates a new,
fascinating field of experimental \cite{48} (FAIR) and theoretical
\cite{21,22,43,46,47} progress.

\section{Conclusions}

It has been the aim of the present paper to shed a light at the
apparent puzzle of hadron/resonance chemical equilibrium, observed
both in elementary and in nucleus-nucleus collisions at $\sqrt{s}
\ge 17$ GeV. To this end we have first revisited the models
developed in the 1970-1980s \cite{1,2,3,4,5} for hadronization as it
occurs in the $e^+e^-$ annihilation reaction. These models develop a
QCD view that offers an evolution which ends, ''naturally'', i.e.
without implausible ad-hoc assumptions, in a decoherent, on-shell
hadron plus resonance ensemble, with a yield order which is
determined by the width of available phase space (represented by an
effective formation temperature) vs. the hadronic mass and spin
spectrum. This situation was initially confronted by Hagedorn
\cite{8,18} within the statistical hadronization model, and more
recently by Becattini, Heinz and collaborators \cite{6,7,20,31}
employing the canonical ensemble. This overall satisfactory
situation in modeling elementary hadronization, which is relatively
simple as no final state interaction modifies the hadronic
equilibrium yield distribution (hadronization and hadronical
freeze-out thus coinciding), suggests an attempt to, likewise,
describe hadronization in A+A collisions. In fact, the situation can
be qualitatively understood with a single additional assumption.
Owing to the high spatial energy density the initial QCD DGLAP
evolution branches might settle, toward the end of the non
perturbative era, into extended configurations of amalgamating
clusters or strings which we have called super-clusters. This
particular non-pertubative symbolic language may, in fact, imply a
general quark gluon plasma formation process. As extended volumes
should decay relatively less constrained by strict local quantum
number requirements \cite{15} but, nevertheless, under phase space
governance, we thus propose that the hadronization output should now
be well described by the grand canonical version of the statistical
model. This implies the observed strangeness saturation systematics
which is well accounted for by the model studies
\cite{11,12,13,14,15,19,20}. Furthermore, also the gradual
transition from canonical (small volume) suppression to grand
canonical saturation, with growing system size, is well documented
by a wealth of experimental \cite{25,32,33,34,35,41} and theoretical
\cite{14,15,26,36,45} studies.

Thus, in returning to the goal, as formulated in the introduction,
we conclude that this ''minimal'' line of argument can, at least
qualitatively, explain the overall hadronization phenomenology. The
puzzling equilibrium distributions may well result from the
stochastic and phase space influences dominating the dynamics, from
late pQCD shower evolution to singlet cluster, or super-cluster,
decay to on-shell hadrons and resonances. From hadronization data
alone we can thus far not derive an argument concerning the
possibility that a genuine partonic equilibrium state precedes the
hadonization mechanisms, at high $\sqrt{s}$ such as explored at
RHIC. We note, however, that other physics observables, e.g.
elliptic flow, support such a more primordial partonic equilibrium
pattern \cite{17}: an important conclusion from the RHIC data.

While insisting on the overall conclusiveness of our ''minimal''
line of argument (not at all a ''deus ex machina'' as argued in
\cite{12}) we observe, finally, that the proposals of super-rapid
equilibration mechanisms, setting in within the instant of parton to
hadron conversion at a high prevailing energy density \cite{21,22}
in A+A collisions, need further investigation. At present they may
be seen to add quantum mechanical detail (relaxation processes with
time scale below hadron size can not be classical) to the overall
''black box'' of quantum mechanical singlet cluster/super-cluster
decay to on-shell hadrons, postulated here. At lower $\sqrt{s}$, in
turn, such process may be dominating the evolution toward hadronic
freeze-out as we have argued in sect. 4.3.


\begin{thebibliography}{99}
\bibitem{1} D. Amati and G. Veneziano, Phys. Lett. B83 (1979) 87
\bibitem{2} B. R. Webber, Nucl. Phys. B238 (1984) 492
\bibitem{3} J. Ellis and K. Geiger, hep-ph/9503349, and Phys. Rev.
D54 (1996) 1967
\bibitem{4} B. Andersson, G. Gustafson, G. Ingelman and T.
Sj\"ostrand, Phys. Rep. 97 (1983) 33
\bibitem{5} R. D. Field and S. Wolfram, Nucl. Phys. B213 (1983) 65;
B. R. Webber, hep-ph/9411384; B. R. Webber, Ann. Rev. Nucl. Part.
Sci. 36 (1986) 253
\bibitem{6} F. Becattini, Nucl. Phys. A702 (2002) 336
\bibitem{7} F. Becattini and G. Passaleva, Eur. Phys. Journ. C23
(2002) 551
\bibitem{8} R. Hagedorn, CERN Yellow Report 71-72 (1971)
\bibitem{9} G. Marchesini, L. Trentadue and G. Veneziano, Nucl. Phys.
B181 (1980) 335
\bibitem{10} T. Ericson and W. Weise, Oxford, Clarendon 1988; W. Weise,
nucl-th/0504087
\bibitem{11} F. Becattini, M. Gazdzicki, A. Ker\"anen, J. Manninen
and R. Stock, Phys. Rev. C69 (2004) 024905
\bibitem{12} A. Andronic, P. Braun-Munzinger and J. Stachel,
nucl-th/0511071, and references therein
\bibitem{13} F. Becattini, J. Manninen and M. Gazdzicki,
hep-ph/0511092
\bibitem{14} J. Cleymans, H. Oeschler, K. Redlich and S. Wheaton,
Eur. Phys. J. A29 (2006) 119; J. Cleymans, M. Stankiewicz and P.
Steinberg, nucl-th/0506027; J. Cleymans, B. Kaempfer, M. Kaneta, S.
Wheaton and N. Xu, Phys. Rev. C71  (2005) 054901
\bibitem{15} A. Tounsi and K. Redlich, J. Phys. G28 (2002) 2095; J.
Rafelski and M. Danos, Phys. Lett. B97 (1980) 279; J. Rafelski and
J. Letessier, J. Phys. G28 (2002) 1819
\bibitem{16} B. M\"uller and J. Rafelski, Phys. Rev. Lett. 48 (1982)
1066
\bibitem{17} M. Gyulassy and L. McLerran, nucl-th/0405013; R. A.
Lacey and A. Taranenko, nucl-ex/0610029
\bibitem{18} R. Hagedorn, Nucl. Phys. B24 (1979) 93; R. Stock, Phys.
Rep. 135 (1986) 259
\bibitem{19} R. Stock, Phys. Lett. B456 (1999) 277
\bibitem{20} U. Heinz, Nucl. Phys. A661 (1999) 140, and talk at this
conference
\bibitem{21} P. Braun-Munzinger, J. Stachel and Ch. Wetterich, Phys.
Lett. B596 (2004) 61
\bibitem{22} C. Greiner, P. Koch-Sternheimer, F. M. Liu I. A.
Shovkovy and H. St\"ocker, J. Phys. G31 (2005) 725, and references
therein
\bibitem{23} F. Karsch, Nucl. Phys. A698 (2002) 199
\bibitem{24} C. A. Salgado und U. Wiedemann, Phys. Rev. D68 (2003)
014008
\bibitem{25} C. Alt et al., NA49 Coll., Phys. Rev. Lett. 94 (2005)
052301
\bibitem{26} C. Hoehne, F. Puehlhofer and R. Stock, Phys. Letters
B640 (2006) 96
\bibitem{27} Z. Fodor and S.D. Katz, J. High Energ. Phys. 203 (2002)
14; ibidem 404 (2004) 50
\bibitem{28} A. K. Wroblewski, Acta Phys. Pol B16 (1985) 379
\bibitem{29} S. V. Afanasiev et al., NA49 Coll., Phys. Rev. C66
(2002) 054902, and Phys. Rev. C69 (2004) 024902
\bibitem{30} S. A. Bass and A. Dumitru, Phys. Rev. C61 (2000) 064909
\bibitem{31} F. Becattini and U. Heinz, Z. Phys. C76 (1997) 269
\bibitem{32} D. Elia et al., NA57 Coll., J. Phys. G31 (2005) 135; F.
Antinori et al., NA57 Coll., J. Phys. G32 (2006) 427; I. Kraus et
al., NA49 Coll. J. Phys. G31 (2005) 147; A. Richard et al., NA49
Coll., ibidem p. 155; H. Caines et at al., STAR Coll., ibidem p. 101
\bibitem{33} L. Ahle et al., E802 Coll., Phys. Rev. C60 (1999)
064901
\bibitem{34} M. Mitrovski et al., NA49 Coll., Proceedings QM06, to
be published in J. Phys. G., nucl-ex/0606004
\bibitem{35} H. Caines, STAR Coll., ibidem, nucl-ex/0608008
\bibitem{36} J. Cleymans, J. Phys. G28 (2002) 1575
\bibitem{37} S. S. Adler et al., PHENIX Coll., Phys. Rev. C69 (2004)
034909
\bibitem{38} I. G. Baerden et al., BRAHMS Coll., Phys. Rev. Lett. 90
(2003) 102301; D. Ouerdane et al., BRAHMS Coll., J. Phys. G30 (2004)
1129
\bibitem{39} D. R\"ohrich, talk at this Conference
\bibitem{40} C. Alt et al., NA49 Coll., Phys. Rev. C73 (2006) 044910
\bibitem{41} G. Adams et al., STAR Coll., nucl-ex/0604014
\bibitem{42} F. Becattini, private communication
\bibitem{43} D. Zschiesche, talk at this Conference
\bibitem{44} P. Braun-Munzinger, J. Cleymans, H. Oeschler and K.
Redlich, Nucl. Phys. A697 (2002) 902
\bibitem{45} J. Cleymans, talk at this Conference
\bibitem{46} R. Stock, J. Phys. G30 (2004) 633
\bibitem{47} V. Koch, A. Majumder and J. Randrup, nucl-th/0509030;
M. Asakawa and C. Nonaka, nucl-th/050991
\bibitem{48} M. Gazdzicki, talk at this Conference
\end{thebibliography}
\end{document}